\RequirePackage{fix-cm}
\documentclass{svjour3}
\usepackage{amssymb}
\usepackage{amsmath}
\usepackage{graphicx}
\usepackage{subfigure}

\begin{document}
\title{Entropic uncertainty in the background of expanding de Sitter space-time}
\author{Zhiming Huang}
\institute{
Zhiming Huang \at
School of Economics and Management, Wuyi University, Jiangmen 529020, China \\
\email{465609785@qq.com}\\
}
\maketitle
\begin{abstract}
We study the dynamics of quantum-memory-assisted entropic uncertainty for a hybrid qutrit-qubit system interacting with fluctuating quantum scalar field in the background of expanding de Sitter space. We firstly derive the master equation that the system evolution obeys. As evolution time goes by, for different initial states, entropic uncertainty develops to different fixed values for different parameter values, whereas entanglement always decays to zero, and there exist monotonous relations between entropic uncertainty, entanglement and various parameters for a fixed initial state, but mixedness behaves differently with entropic uncertainty and entanglement. Further it is found that the entropic uncertainty closely associated with the entanglement and mixedness. In addition, it is shown that the entropic uncertainty can be manipulated effectively via the weak measurement reversal. Our study would give some useful insights about the behavior characteristics of high dimensional quantum system in expanding de Sitter space-time, and may be useful to the tasks of quantum information processing of curved space-time since the uncertainty principle plays vital role in quantum information science and technology.
\end{abstract}
\section{Introduction}
The uncertainty principle \cite{Sen2014} as a vital feature of quantum physics restricts our ability to simultaneously predict the precise outcomes of two incompatible observables.
Since Heisenberg \cite{Heisenberg1927} proposed the first expression of the uncertainty principle versus position and momentum, afterwards various different forms of uncertainty relation are introduced.
Some authors \cite{Kennard1927,Robertson1929} generalized to the standard deviation form which is state-dependent and is not the optimal. To avoid the drawback, some authors \cite{Deutsch1983,Kraus1987,Maassen1988,Bialynicki2006} introduced entropic uncertainty relation, which lower bound is state independent. Recently, a stronger entropic uncertainty relation called quantum-memory-assisted entropic uncertainty is proposed  \cite{Renes2009,Berta2010}. The uncertainty relation has many potential applications, such as probing quantum correlation \cite{Hu2013a,Pati2012}, quantum speed limit \cite{Mondal2016,Pires2016} and quantum key distribution \cite{Tomamichel2012,Coles2014}. In addition, Quantum entanglement is the most fascinating characteristic of quantum mechanics and plays significant role in various quantum information processing tasks \cite{Horodecki2009,Nilsen2000}. The entropic uncertainty is closely related with quantum entanglement and is as a novel witness of quantum entanglement \cite{Hu2012a,Zou2014}.

In the real world, quantum systems unavoidably are affected by surrounding environment, for instance, vacuum fluctuation of quantum field has a great influence on quantum system under it. Actually, vacuum fluctuation is a natural consequence of the uncertainty principle. Recently, Huang et al. \cite{Huang2018a,Huang2019} investigated the entropic uncertainty relation affected by the vacuum fluctuation of scalar field and electromagnetic field. Besides, the curvature of the space-time background also affects the behaviors of open quantum systems. The entanglement dynamics of atoms coupling with a fluctuating scalar field in de Sitter space-time was investigated in Ref. \cite{Tian2014,Huang2017a}. Further some authors discussed the behaviors of quantum correlation for atoms immersing in a fluctuating scalar field in de Sitter space-time \cite{Huang2017b,Feng2018}. Some authors explored the quantum estimation in de Sitter space-time \cite{Huang2018b,Huang2018}.

According to the current observations and the theory of inflation, de Sitter space-time approximates the geometries of our universe in the far past and the far future.
Quantum fluctuation is assumed to start in a Bunch-Davies vacuum at infinite past \cite{Feng2018}, which is dubious since the field modes below the Planck scale are inaccessible. The de Sitter invariant Bunch-Davies
vacuum can be extended to the $\alpha$-vacuum, which can be interpreted as a squeezed state over Bunch-Davies vacuum \cite{Huang2018} and heavily constrain the measurement uncertainty. In this paper, we study the entropic uncertainty of $\alpha$-vacua in the background of de Sitter space-time. We would study the behaviors of entropic uncertainty in the framework of open quantum system that a qutrit-qubit system couples with fluctuating quantum scalar field in de Sitter space-time. Further we investigate the relations of quantum-memory-assisted entropic uncertainty, entanglement, mixedness, and steering entropic uncertainty with weak measurement reversal under $\alpha$-vacua fluctuation. Our study would elucidate the behaviors and relations of entropic uncertainty, entanglement, mixedness, and how to adjust the entropic uncertainty affected by fluctuating $\alpha$-vacuum in de Sitter space-time. Our exploration may be helpful for us understanding entropic uncertainty dynamics of high dimensional quantum system in the expanding curved space-time.

In the following, we firstly introduce the quantum-memory-assisted entropic uncertainty in term of the game model in de Sitter space-time, entanglement measure, mixedness and weak measurement reversal, and the evolution physical model. And then we derive the master equation that describes the system evolution, and discuss the dynamical behaviors of the entropic uncertainty affected by $\alpha$-vacuum fluctuation in de Sitter universe and how to adjust the behaviors of the entropic uncertainty via weak measurement reversal. Finally, we present a conclusion of this paper.

\section{Preliminaries}\label{S2}
The entropic uncertainty relation can be described as a guessing game between two player Alice and Bob. Assuming Bob prepares an entangled qutrit-qubit state $\rho_{UV}$ in flat Minkowski
space-time and sends a qutrit $U$ to Alice and keeps the qubit $V$ as a quantum memory for himself. After a certain moment, the qubit $V$ freely falls toward the de Sitter space-time and hovers near the event horizon of the de Sitter space, whereas the qutrit $U$ stays at the flat Minkowski space-time. Subsequently,
Alice executed one of the two measurements $O_1$ and $O_2$ on the qutrit $U$ and informs her choice of measurement to Bob. Based on his quantum memory $V$, Bob needs to guess the measurement
outcome, and if he guesses the outcome correctly, he win the game.
Quantum-memory-assisted entropic uncertainty relation for two incompatible observables $O_1$ and $O_2$ reads as \cite{Renes2009,Berta2010}
\begin{align}
S(O_1|V)+S(O_2|V)\geq S(U|V)+\log_2 \frac{1}{c},\label{EUR}
\end{align}
where $S(U|V)=S(\rho_{UV})-S(\rho_V)$ is the conditional von Neumann entropy with $S(\rho)=-\text{Tr}(\rho\log_2\rho)$, $c=\max_{ij}|\langle \phi_i|\varphi_j\rangle|^{2}$ with $|\phi_i\rangle$ and $|\varphi_i\rangle$ being the eigenstates of the observable $O_1$ and $O_2$. After qubit $U$ is measured by $O_1$, the post-measurement state becomes $\rho_{O_1 V}=\sum_i (|\phi_i\rangle\langle \phi_i|\otimes I)\rho_{UV}(|\phi_i\rangle\langle \phi_i|\otimes I)$. If measured qutrit $U$ and qubit $V$ are entangled, Bob\rq s uncertainty about Alice\rq s measurement outcome would be reduced. Here we use the notation $L$ to denote the left side of the uncertainty relation, i.e., $L\equiv S(O_1|V)+S(O_2|V)$, and $R$ denotes the right side of the uncertainty relation, i.e., $R\equiv S(U|V)+\log_2 \frac{1}{c}$.

To discuss the relationship between the entropic uncertainty and the entanglement of the qutrit-qubit state in the expanding de Sitter space-time, we employ the negativity as the measure of entanglement, which is defined as \cite{Vidal2002}
\begin{align}
N={\|\rho^{T_{U}}\|-1},\label{ne2}
\end{align}
where $\|\rho\|=\text{Tr}(\sqrt{\rho^{\dag}\rho})$ is the trace norm, $\rho^{T_{U}}$ denotes the partial transpose of the state with respect to the subsystem $U$.

For analyzing the relationship between the entropic uncertainty and the mixedness, we introduce the mixedness here. For a general quantum state $\rho$, the state is pure when $\text{Tr}(\rho^{2})=1$, and is mixed if $\text{Tr}(\rho^{2})<1$. Thus the mixedness can be defined as \cite{Peters2004}
\begin{align}
X=\frac{d}{d-1}[1-\text{Tr}(\rho^{2})],\label{mixn}
\end{align}
where $d$ is the dimension of state $\rho$.

In order to discuss the manipulation of the uncertainty relation, we introduce the quantum weak measurement reversal \cite{Sun2010}, which with strength $p$ ($0\leq p< 1$) is written as
\begin{align}
M=\sqrt{1-p}|0\rangle\langle0|+|1\rangle\langle1|=
\left[
\begin{array}{cc}
  \sqrt{1-p} & 0 \\
  0 & 1 \\
\end{array}
\right].\label{WMR}
\end{align}
After the weak measurement reversal is applied to a bipartite state $\rho$, the post-selection state of the bipartite state becomes
\begin{align}
\rho^\prime=\frac{(I\otimes M) \rho(I\otimes M)^{\dagger}}{\mathrm{Tr}[(I\otimes M) \rho (I\otimes M)^{\dagger}]}.
\end{align}

In our model, we consider the qubit system weakly couples with a bath of fluctuating quantum scalar field in de Sitter space-time.
The whole Hamiltonian of such a system takes the form
\begin{align}
H=H_{S}+H_{F}+H_{I}.
\end{align}
 $H_{S}=\frac{\omega}{2}\sigma_{3}$ is the Hamiltonian of the system, where $\sigma_{i}$ $(i=1,2,3)$ are the Pauli matrices, and $\omega$ is the energy level spacing. $H_{F}$ is the Hamiltonian of the scalar field. $H_I=\sigma_{2}\Phi(x)$ is the interaction Hamiltonian between the system and the scalar field.

We assume the initial state of the total system takes the form $\rho_{tot}(0)=\rho(0)\otimes|v\rangle\langle v|$, where $\rho(0)$ is the initial state of the quantum system and $|v\rangle$ is the vacuum state of the external field. Under the Born-Markov approximation, the system evolution in the proper time $\tau$ of the system can be described with the Kossakowski-Lindblad master equation \cite{Gorini1976,Lindblad1976,Breuer2002}
\begin{align}
\frac{\partial\rho(\tau)}{\partial \tau}=-\rm{i}[H_{\text{eff}},\rho(\tau)]+\mathcal{L}[\rho(\tau)],\label{m1}
\end{align}
where
\begin{align}
H_{\text{eff}}=\frac{1}{2}\Omega\sigma_{3}=\frac{1}{2}\{\omega+\frac{i}{2}[\mathcal{K}(-\omega)-\mathcal{K}(\omega)]\}\sigma_{3},
\end{align}
and
\begin{align}
\mathcal{L}[\rho]=\frac{1}{2}\sum_{i,j=1}^{3}S_{ij}[2\sigma_{j}\rho\sigma_{i}-\sigma_{i}\sigma_{j}\rho-\rho\sigma_{i}\sigma_{j}].\label{m2}
\end{align}
$S_{ij}$ can be written explicitly as
\begin{align}
S_{ij}= A\delta_{ij} -iB\epsilon_{ijk}\,\delta_{3k} -A\delta_{3i}\,\delta_{3j} ,\label{CC1}
\end{align}
where
\begin{align}
A={\frac{1}{4}}\,[\,{\cal G}(\omega) +{\cal G}(-\omega)] ,\quad B ={\frac{1}{4}}\,[\,{\cal G}(\omega) -{\cal G}(-\omega)].\label{AB1}
\end{align}
$\mathcal{G}(\lambda)$ and $\mathcal{K}(\lambda)$ denote the Fourier and Hilbert transforms of field correlation function $G(\tau-\tau')$ respectively, defined as
\begin{align}
&\mathcal{G}(\lambda)=\int_{-\infty}^{\infty}d\Delta \tau e^{\rm{i} \lambda \Delta \tau}G(\Delta \tau),\label{ft}\\
&\mathcal{K}(\lambda)=\frac{P}{\pi \rm{i}}\int_{-\infty}^{\infty}d\omega\frac{\mathcal{G}(\omega)}{\omega-\lambda}.
\end{align}
 Note that $c = \hbar = 1$ throughout this article.

\section{Dynamics of entropic uncertainty}\label{S3}
Considering a freely falling qubit in de Sitter space weakly couples with a massless scalar field. Choosing the global coordinate system $({t},\chi,\theta,\varphi)$, the qubit is comoving with the expansion, then the line element of de Sitter space-time can be written as
\begin{align}
ds^2=dt^2-\ell^2\cosh^2(t/\ell)[d\chi^2+\sin\chi^2(d\theta^2+\sin^2\theta d\varphi^2)], \label{dsl}
\end{align}
where curvature radius $\ell=3^{1/2}\Lambda^{-1/2}$, with $\Lambda$ being the cosmological constant.  $\ell$ is associated with the Gibbons-Hawking temperature $T =\frac{1}{2\pi \ell}$. When we choose the de Sitter-invariant Bunch-Davies vacuum state $|0\rangle$ as the state of the conformally coupled massless scalar field, the corresponding massless conformally Wightman function for the qubit system is
\begin{eqnarray}
G^+(x(\tau),x(\tau'))=\langle0|\Phi(x(\tau))\Phi(x(\tau'))|0\rangle=-\frac{1}{16\pi^2\ell^2\sinh^2({\frac{\tau-\tau'}{2\ell}}-i\varepsilon)},\label{bdw}
\end{eqnarray}
which fulfills KMS condition \cite{Kubo1957,Martin1959,Haag1967}. Further one can generalize the Bunch-Davies vacuum to de Sitter invariant vacua called $\alpha$-vacua $|\alpha\rangle$, which are believed to play an important role in understanding the trans-Planckian physics of early universe. $\alpha$-vacua can be generated by a squeezing operator $\hat{S}(\alpha)$ \cite{Barnett1997}, i.e., $|\alpha\rangle=\hat{S}(\alpha)|0\rangle$, where $\alpha<0$ is real when we adopt CPT invariant $\alpha$-vacua. The Wightman function for the scalar field of $\alpha$-vacua can be written with the Wightman function $G^+(x,y)$ of Bunch-Davies vacuum as
\begin{align}
G^+_\alpha(x,y)=\frac{1}{1-e^{2\alpha}}\Big[G^+(x,y)+e^{2\alpha}G^+(y,x)-e^{\alpha}\big(G^+(x,y_A)+ G^+(x_A,y)\big)\Big],\label{nbdw}
\end{align}
where $x_A$ is the antipodal point of $x$. When $\alpha\rightarrow -\infty$, $G^+_\alpha(x,y)$ reduces to the Wightman function $G^+(x,y)$ of Bunch-Davies vacuum.
According to the relation $ G^+(x,y_A)=G^+(x_A,y)=G^+(\tau-i\pi)$ \cite{Bousso2002} and Eq. (\ref{bdw}), we can obtain the Fourier transformation (\ref{ft}) of correlation function $G^+_\alpha(x,y)$ (\ref{nbdw})
\begin{align}
\mathcal{G}_\alpha=\frac{\lambda(1+e^{\alpha-\pi\lambda})^2}{2\pi(1-e^{-\frac{\lambda}{T}})(1-e^{2\alpha})}
\end{align}
for a conformally coupling massless scalar field in general $\alpha-$vacua. According to  Eq. (\ref{AB1}), the coefficients of $S_{ij}$ (\ref{CC1}) can be obtained
\begin{align}
&A=\frac{\omega  [(e^{\alpha -\pi  \omega }+1)^2+(e^{\alpha +\pi  \omega }+1)^2 e^{-\frac{\omega }{T}}]}{8 \pi  (e^{2 \alpha }-1) (e^{-\frac{\omega }{T}}-1)},\nonumber\\
&B=\frac{\omega  [(e^{\alpha -\pi  \omega }+1)^2-(e^{\alpha +\pi  \omega }+1)^2 e^{-\frac{\omega }{T}}]}{8 \pi  (e^{2 \alpha }-1) (e^{-\frac{\omega }{T}}-1)}.
\end{align}

Now, we consider to solve the master equation (\ref{m1}) for an initial qutrit-qubit state. Note that the master equation for qubit (\ref{m1}) can still work, because here we assume only the qubit system $V$ is affected by $\alpha-$vacua fluctuation and the qutrit system $U$ is isolated from the external environment. Thus we only need to slightly change the master equation (\ref{m1}) to the following the form:
\begin{align}
\frac{\partial\rho(\tau)}{\partial \tau}=-\rm{i}[H'_{\text{eff}},\rho(\tau)]+\mathcal{L'}[\rho(\tau)],\label{m3}
\end{align}
where $ H'_{\text{eff}}=\frac{1}{2}\Omega(I\otimes\sigma_{3})=\frac{1}{2}\{\omega+\frac{i}{2}[\mathcal{K}(-\omega)-\mathcal{K}(\omega)]\}(I\otimes\sigma_{3}), $
$\mathcal{L'}[\rho]=\frac{1}{2}\sum_{i,j=1}^{3}S_{ij}[2(I\otimes\sigma_{j})\rho(I\otimes\sigma_{i})-(I\otimes\sigma_{i})(I\otimes\sigma_{j})\rho-\rho(I\otimes\sigma_{i})(I\otimes\sigma_{j})]$, with $I$ being the $3\times 3$ identity operator.
Any qutrit-qubit state can be written as
\begin{align}
\rho =\left(
\begin{array}{cccccc}
 a_{1,1}(t) & a_{1,2}(t) & a_{1,3}(t) & a_{1,4}(t) & a_{1,5}(t) & a_{1,6}(t) \\
 a_{2,1}(t) & a_{2,2}(t) & a_{2,3}(t) & a_{2,4}(t) & a_{2,5}(t) & a_{2,6}(t) \\
 a_{3,1}(t) & a_{3,2}(t) & a_{3,3}(t) & a_{3,4}(t) & a_{3,5}(t) & a_{3,6}(t) \\
 a_{4,1}(t) & a_{4,2}(t) & a_{4,3}(t) & a_{4,4}(t) & a_{4,5}(t) & a_{4,6}(t) \\
 a_{5,1}(t) & a_{5,2}(t) & a_{5,3}(t) & a_{5,4}(t) & a_{5,5}(t) & a_{5,6}(t) \\
 a_{6,1}(t) & a_{6,2}(t) & a_{6,3}(t) & a_{6,4}(t) & a_{6,5}(t) & a_{6,6}(t) \\
\end{array}
\right). \label{tes}
\end{align}
We consider the qutrit-qubit system $U$ and $V$ initially share the following state
\begin{align}
\rho(0) =\left(
\begin{array}{cccccc}
 \frac{f}{2} & 0 & 0 & 0 & 0 & \frac{f}{2} \\
 0 & \frac{1}{2} (1-2 f) & 0 & 0 & \frac{1}{2} (1-2 f) & 0 \\
 0 & 0 & \frac{f}{2} & 0 & 0 & 0 \\
 0 & 0 & 0 & \frac{f}{2} & 0 & 0 \\
 0 & \frac{1}{2} (1-2 f) & 0 & 0 & \frac{1}{2} (1-2 f) & 0 \\
 \frac{f}{2} & 0 & 0 & 0 & 0 & \frac{f}{2} \\
\end{array}
\right), \label{s0}
\end{align}
where $0\leq f \leq \frac{1}{3}$, $f=0$ corresponds to the maximally entangled state, and $f=\frac{1}{3}$ corresponds to the separable state.

By substituting Eq. (\ref{tes}) into Eq. (\ref{m3}), and according to the initial state \label{s0}, we can get the evolution state
\begin{align}
\rho_{UV}=\left(
\begin{array}{cccccc}
 Q_1 & 0 & 0 & 0 & 0 & Q_7 \\
 0 & Q_2 & 0 & 0 & Q_8 & 0 \\
 0 & 0 &Q_3 & 0 & 0 & 0 \\
 0 & 0 & 0 & Q_4& 0 & 0 \\
 0 & Q_8 & 0 & 0 & Q_5 & 0 \\
Q_7 & 0 & 0 & 0 & 0 & Q_6 \\
\end{array}
\right),\label{fs}
\end{align}
where $Q_1= \frac{e^{-4 A \tau} [3 f A-A+B-(A-B) e^{4 A \tau} (f-1)-B f]}{4 A}$, $Q_2=\frac{e^{-4 A \tau} [-3 f A+A-B-(A+B) e^{4 A \tau} (f-1)+B f]}{4 A}$, $Q_3=  \frac{(A-B+Be^{-4 A \tau}) f}{2 A}$, $Q_4= \frac{(A-B e^{-4 A \tau}+B) f}{2 A}$, $Q_5=\frac{e^{-4 A \tau} [-3 f A+A+B-(A-B) e^{4 A t} (f-1)-B f]}{4 A}$, $Q_6=\frac{e^{-4 A \tau} [3 f A-A-B-(A+B) e^{4 A t} (f-1)+B f]}{4 A}$, $Q_7=\frac{1}{2} e^{-2 A \tau} f$ and $Q_8=\frac{1}{2} e^{-2 A \tau} (1-2 f)$.
Considering two Pauli observables $\sigma_z=|0\rangle\langle0|-|2\rangle\langle2|$ and $\sigma_x=\frac{1}{\sqrt{2}}(|0\rangle\langle1|+|1\rangle\langle0|+|1\rangle\langle2|+|2\rangle\langle1|)$ as two incompatible measurements performed on the subsystem $U$, in this case, the parameter $c$ in Eq. (\ref{EUR}) is exactly equal to 1/2, and the post-measurement state can be obtained
\begin{align}
&\rho_{\sigma_z V}=\left(
\begin{array}{cccccc}
Q_1 & 0 & 0 & 0 & 0 & 0 \\
 0 & Q_2 & 0 & 0 & 0 & 0 \\
 0 & 0 & Q_3 & 0 & 0 & 0 \\
 0 & 0 & 0 & Q_4 & 0 & 0 \\
 0 & 0 & 0 & 0 & Q_5 & 0 \\
 0 & 0 & 0 & 0 & 0 &Q_6 \\
\end{array}
\right), \nonumber\\
&\rho_{\sigma_x V}=\frac{1}{4}\left(
\begin{array}{cccccc}
F_1& F_2 & 0 & 0 &F_3 & F_4 \\
F_2 & F_5 & 0 & 0 & F_4 & F_6 \\
 0 & 0 & F_7 & F_8 & 0 & 0 \\
 0 & 0 & F_8 & F_9 & 0 & 0 \\
F_3 & F_4 & 0 & 0 &F_1 & F_2 \\
F_4 & F_6 & 0 & 0 & F_2 & F_{5} \\
\end{array}
\right),
\end{align}
where $F_1= \frac{e^{-4 A \tau} [(B-A) e^{4 A t\tau}-B] (f-3)}{4 A} $, $F_2=\frac{1}{4} e^{-2 A \tau} (f-1)$, $F_3= \frac{e^{-4 A \tau} [e^{4 A \tau} (A-B)+B] (3 f-1)}{4 A}$, $F_4=-\frac{3}{4} e^{-2 A \tau} (f-1)$, $F_5=\frac{e^{-4 A \tau} [B-(A+B) e^{4 A \tau}] (f-3)}{4 A}$, $F_6=\frac{e^{-4 A \tau}[(A+B) e^{4 A \tau}-B](3 f-1)}{4 A}$, $F_7=\frac{e^{-4 A \tau} [e^{4 A \tau} (A-B)+B](f+1)}{2 A}$, $F_8=-\frac{1}{2} e^{-2 A \tau} (f-1)$ and $F_9=\frac{e^{-4 A \tau} [(A+B) e^{4 A \tau}-B](f+1)}{2 A}$.
The reduced density matrix of qubit $V$ can be easily obtained by tracing over qutrit $U$
\begin{align}
\rho_{V}=\left(
\begin{array}{cc}
 \frac{A+B (-1+e^{-4 A t})}{2 A} & 0 \\
 0 & \frac{A-B e^{-4 A t}+B}{2 A} \\
\end{array}
\right),
\end{align}
which is independent on parameter $f$.
Because directly calculating the analytic expression of entropic uncertainty takes long time and is complicated, so we mainly employ the numerical method to obtain the results of entropic uncertainty in the following discussion. According to Eq. (\ref{mixn}), we can obtain the mixedness $X=-\frac{3 e^{-8 A t} \{A^2 [(f (3 f-2)-3) e^{8 A t}+2 (f (5 f-4)+1) e^{4 A t}+(1-3 f)^2]+B^2 (f (3 f-2)+1) (e^{4 A t}-1)^2\}}{10 A^2}$. The actual expression of negativity is neglected here since it is long.

Let us firstly consider the equilibrium state. Let $\tau\rightarrow \infty$ in Eq. (\ref{fs}), the equilibrium state can be obtained
\begin{align}
\rho_{\infty}=\left(
\begin{array}{cccccc}
 K_1 & 0 & 0 & 0 & 0 & 0 \\
 0 & K_2 & 0 & 0 & 0 & 0 \\
 0 & 0 & K_3 & 0 & 0 & 0 \\
 0 & 0 & 0 & K_4 & 0 & 0 \\
 0 & 0 & 0 & 0 & K_1 & 0 \\
 0 & 0 & 0 & 0 & 0 & K_2 \\
\end{array}
\right),
\end{align}
where $K_1=-\frac{(A-B) (f-1)}{4 A}$, $K_2=-\frac{(A+B) (f-1)}{4 A}$, $K_3=\frac{(A-B) f}{2 A}$ and $K_4=\frac{(A+B) f}{2 A}$. We know that the equilibrium state is dependent on the initial state. For the state, the negativity vanishes, which means that entanglement only can last for a while. According numerical analysis, it is found that for different initial state, the entropic uncertainty $L$ and $R$ have different fixed values which is independent on the other parameters. But the mixedness $X=\frac{A^2 (-9 f^2+6 f+9)-3 B^2 (3 f^2-2 f+1)}{10 A^2}$, which variations are dependent on various paraments. Note that the equilibrium state is not a thermal state because the Wightman function (\ref{nbdw}) of the $\alpha$-vacua scalar field does not satisfy the KMS condition \cite{Kubo1957,Martin1959,Haag1967}, neither does the reduced state by tracing over the qutrit $U$
\begin{align}
\rho_{\infty V}=\left(
\begin{array}{cc}
 \frac{A-B}{2 A} & 0 \\
 0 & \frac{A+B}{2 A} \\
\end{array}
\right). \label{es2}
\end{align}
But when $\alpha\rightarrow -\infty$, the state (\ref{es2}) reduces to the Bunch-Davies vacuum state, which is equal to the thermal state $e^{-H_S/T}/\text{Tr}(e^{-H_S/T})$. This is a manifestation of thermalization phenomena of de Sitter space-time, which is in agreement with the result of Ref. \cite{Yu2011}.
\begin{figure}
\centering
\subfigure[]{\includegraphics[height=2.5cm,width=3.9cm]{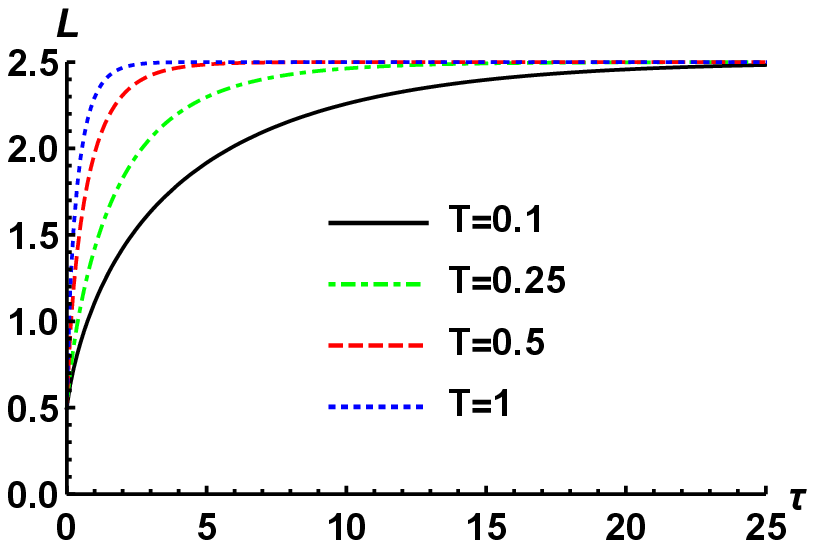}}
\subfigure[]{\includegraphics[height=2.5cm,width=3.9cm]{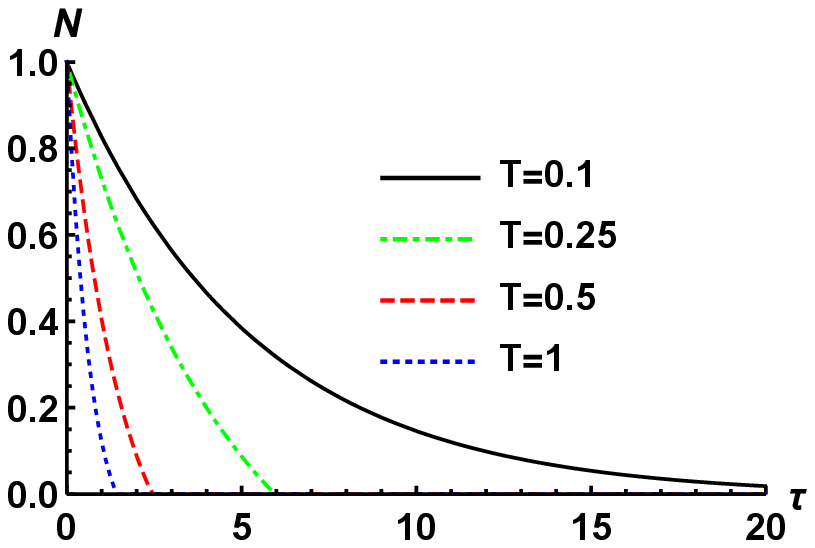}}
\subfigure[]{\includegraphics[height=2.5cm,width=3.9cm]{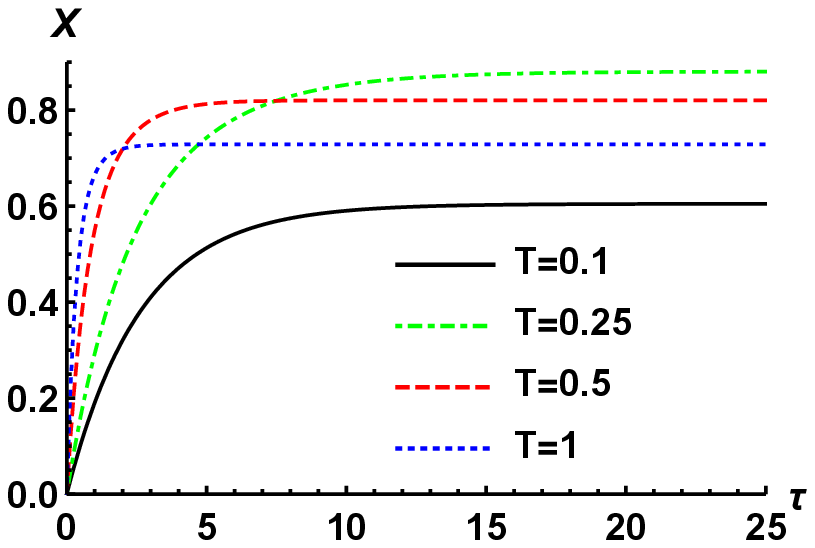}}
\caption{\label{figure1} Entropic uncertainty $L$ (a), entanglement (b) and mixedness (c) as function of $\tau$ for different temperatures with $f=0$, $\alpha=-1$ and $\omega=1$.}
\end{figure}
\begin{figure}
\centering
\includegraphics[height=2.5cm,width=3.9cm]{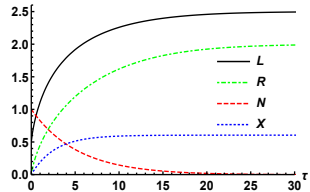}
\caption{\label{figure2} Entropic uncertainty $L$ and $R$, entanglement and mixedness as function of $\tau$ for different temperatures with $f=0$, $T=0.1$, $\alpha=-1$ and $\omega=1$.}
\end{figure}
\begin{figure}
\centering
\subfigure[]{\includegraphics[height=2.5cm,width=3.9cm]{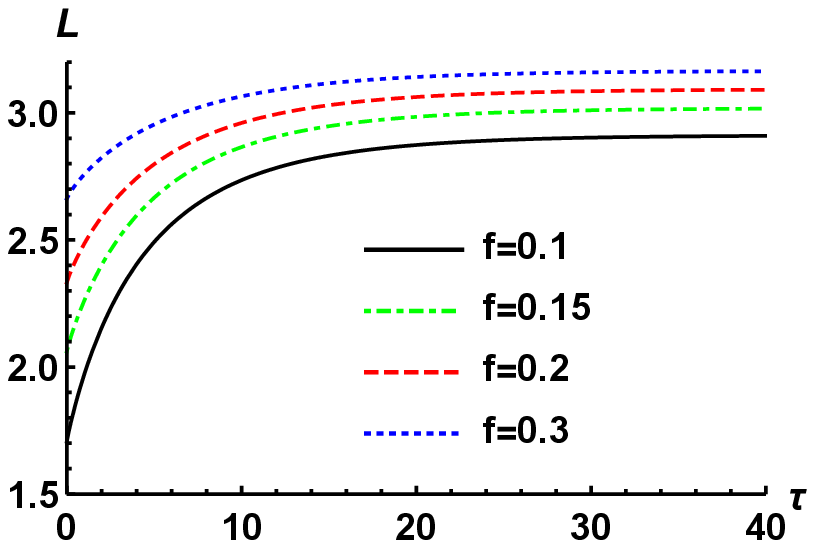}}
\subfigure[]{\includegraphics[height=2.5cm,width=3.9cm]{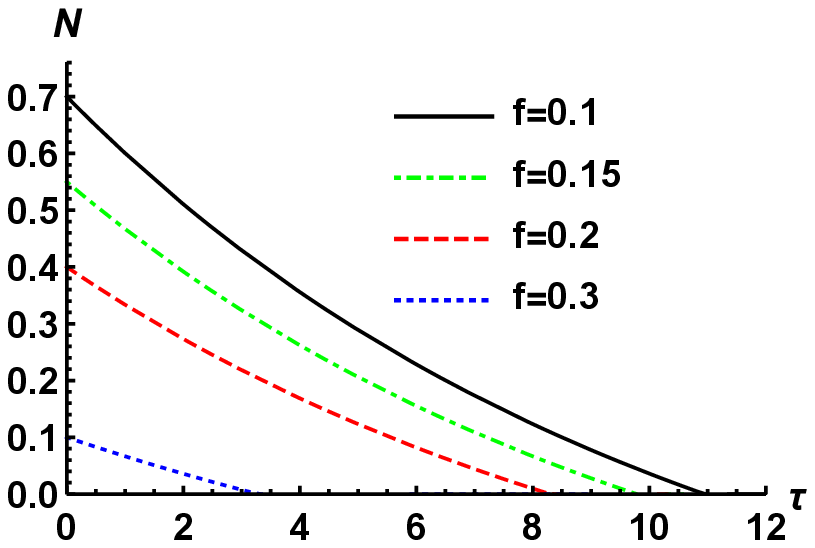}}
\subfigure[]{\includegraphics[height=2.5cm,width=3.9cm]{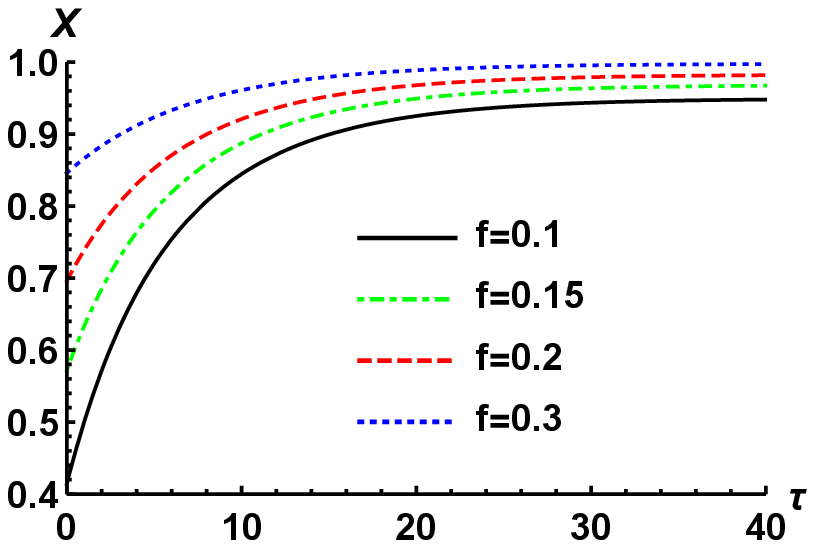}}
\caption{\label{figure3} Entropic uncertainty $L$ (a), entanglement (b) and mixedness (c) as function of $\tau$ for different $f$ values with $T=0.2$, $\alpha=-1$ and $\omega=0.1$.}
\end{figure}
\begin{figure}
\centering
\includegraphics[height=2.5cm,width=3.9cm]{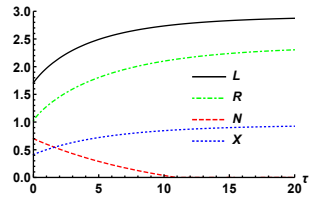}
\caption{\label{figure4} Entropic uncertainty $L$ and $R$, entanglement and mixedness as function of $\tau$ with $f=0.1$, $T=0.2$, $\alpha=-1$ and $\omega=0.1$.}
\end{figure}

Now we consider the general case of Eq. (\ref{fs}). Fig. \ref{figure1} shows the evolution of entropic uncertainty $L$, entanglement and mixedness for different temperatures with the fixed values of other parameters. From Fig. \ref{figure1}, we can see that the entropic uncertainty converges to a constant value 2.5 for different temperatures with evolution time, while entanglement decays to zero with evolution time. Mixedness develops to different stable values for different temperatures. Entropic uncertainty increases as temperature grows for a short time, while entanglement is the opposite, but mixedness does not exist simple monotonic relation with temperature. Higher temperature, entropic uncertainty and mixedness develops to the stable value more rapidly. From Fig. \ref{figure2}, it can be seen that mixedness is positively related with the entropic uncertainty, and entanglement is anti-correlated with entropic uncertainty. In addition, the entropic uncertainty $L$ and $R$ always satisfy the uncertainty relation (\ref{EUR}).

The variations of entropic uncertainty $L$, entanglement and mixedness for different $\alpha$ values ($\alpha\rightarrow-\infty$, $\alpha=-2$, $\alpha=-1$, $\alpha=-0.5$) with the fixed values of other parameters ($f=0$, $T=0.2$ and $\omega=1$) are similar with Fig. \ref{figure1}, but the mixedness similar with the entropic uncertainty exists simple monotonic relation with $\alpha$ parameter.
For simplicity, we ignore the specific figures here.
The variations of entropic uncertainty $L$, entanglement and mixedness for different $\omega$ values ($\omega=0.1$, $\omega=1$, $\omega=2$, $\omega=3$) with the fixed values of other parameters ($f=0$, $T=0.2$ and $\alpha=-1$) are also similar with Fig. \ref{figure1}, and the relations of entropic uncertainty $L$, entanglement and mixedness with the fixed values of various parameters ($f=0$, $\omega=0.1$, $T=0.2$ and $\alpha=-1$) are similar with Fig. \ref{figure2}.

From Fig. \ref{figure3} and Fig. \ref{figure4}, it can be observed that the variations and relations of entropic uncertainty, entanglement and mixedness are similar with Fig. \ref{figure1} and Fig. \ref{figure2} except that the entropic uncertainty and mixedness evolve to different stable values for different initial states, and mixedness decreases as the entanglement increases (the value of $f$ decreases).
Note that these discussed results are consistent with the above analytical results about the equilibrium state.

From above discussions, we know that entanglement exists a certain relation with entropic uncertainty, whereas mixedness has close relation with entropic uncertainty, that is the mixedness can reveal the essence
of the entropic uncertainty better than the entanglement. Temperature originated from space-time curvature and vacuum fluctuation resulted from the uncertainty principle can lead to increase of mixedness and reduction of entanglement, that is temperature and vacuum fluctuation can generate decoherent effect and make an evolution state become more mixed, which results in the observing inaccuracy of quantum system, and naturally brings about an increase of entropic uncertainty.
Further the state mixedness and entropic uncertainty would be helpful to detect thermal effect of curved space-time and quantum fluctuation effect.

Now, consider controlling the entropic uncertainty through weak measurement reversal. When the initial state is executed the weak measurement reversal (\ref{WMR}), the post-measurement state becomes
\begin{align}
\left(
\begin{array}{cccccc}
 f \left(1+\frac{1}{p-2}\right) & 0 & 0 & 0 & 0 & -\frac{f \sqrt{1-p}}{p-2} \\
 0 & \frac{2 f-1}{p-2} & 0 & 0 & \frac{(2 f-1) \sqrt{1-p}}{p-2} & 0 \\
 0 & 0 & f \left(1+\frac{1}{p-2}\right) & 0 & 0 & 0 \\
 0 & 0 & 0 & -\frac{f}{p-2} & 0 & 0 \\
 0 & \frac{(2 f-1) \sqrt{1-p}}{p-2} & 0 & 0 & -\frac{(2 f-1) (p-1)}{p-2} & 0 \\
 -\frac{f \sqrt{1-p}}{p-2} & 0 & 0 & 0 & 0 & -\frac{f}{p-2} \\
\end{array}
\right).
\end{align}
\begin{figure}
\centering
\includegraphics[height=2.5cm,width=3.9cm]{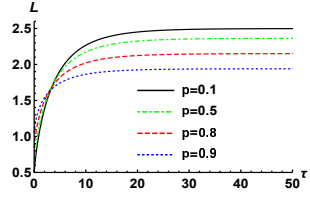}
\caption{\label{figure5} Entropic uncertainty $L$ as function of $\tau$ for different strengths with $f=0$, $T=0.2$, $\alpha=-1$ and $\omega=0.1$.}
\end{figure}
After evolution, we can obtain the evolution state corresponding to post-measurement state. Here we neglect it for simplicity. With the same computing equations and analytical method, one can obtain the behaviors of entropic uncertainty $L$ of the evolution state.
From Fig. \ref{figure5}, it can be seen that the weak measurement reversal can effectively reduce the uncertainty. As evolution time grows, the entropic uncertainty is steered to the different stable values with different strengths of weak measurement reversal.
\section{Conclusion}\label{S4}
We have studied the behaviors and relations of quantum-memory-assisted entropic uncertainty, entanglement and mixedness for qutrit-qubit system weakly coupling with a bath of fluctuating scalar field in the background of expanding de Sitter space. The master equation that governs the system evolution is derived. For different initial states, entropic uncertainty and mixedness develop to different stable values. For different values of other parameters, entropic uncertainty always evolves to a fixed stable, while entanglement always decays to zero, and entropic uncertainty and entanglement exists monotonous relation with various parameters for a short time, but mixedness is not the cases. Besides, it is found that the entropic uncertainty, entanglement and mixedness exists closely relations. In addition, it is shown that the weak measurement reversal can effectively adjust the entropic uncertainty. Our exploration would threw light on our understanding about the quantum nature of high dimensional quantum system in expanding curved space-time and may be helpful for quantum information processing under curved space-time.


\begin{thebibliography}{00}
\bibitem{Sen2014}D. Sen: The uncertainty relations in quantum mechanics. Curr. Sci. 107, 203 (2014)
\bibitem{Heisenberg1927}W. Heisenberg: \"{U}ber den anschaulichen Inhalt der quantentheoretischen Kinematik und Mechanik. Z. Phys. 43, 172 (1927)

\bibitem{Kennard1927}E.H. Kennard: Zur Quantenmechanik einfacher Bewegungstypen. Z. Phys. 44, 326 (1927)
\bibitem{Robertson1929}H.P. Robertson: The uncertainty principle. Phys. Rev. 34, 163 (1929)

\bibitem{Deutsch1983}D. Deutsch: Uncertainty in quantum measurements.  Phys. Rev. Lett. 50, 631 (1983)
\bibitem{Kraus1987}K. Kraus: Complementary observables and uncertainty relations. Phys. Rev. D 35, 3070 (1987)
\bibitem{Maassen1988}H. Maassen, J.B.M. Uffnk: Generalized entropic uncertainty relations. Phys. Rev. Lett. 60, 1103 (1988)
\bibitem{Bialynicki2006}I. Bialynicki-Birula: Formulation of the uncertainty relations in terms of the R\'{e}nyi entropies. Phys. Rev. A 74, 052101 (2006)

\bibitem{Renes2009}J.M. Renes, J.C. Boileau: Conjectured strong complementary information tradeoff. Phys. Rev. Lett. 103, 020402 (2009)
\bibitem{Berta2010}M.Berta, M. Christandl, R. Colbeck, J.M. Renes, R. Renner: The uncertainty principle in the presence of quantum memory. Nat. Phys. 6, 659 (2010)

\bibitem{Hu2013a}M.L. Hu, H. Fan: Competition between quantum correlations in the quantum-memory-assisted entropic uncertainty relation. Phys. Rev. A 87, 022314 (2013)
\bibitem{Pati2012}A.K. Pati, et al.: Quantum discord and classical correlation can tighten the uncertainty principle in the presence of quantum memory. Phys. Rev. A 86, 042105 (2012)

\bibitem{Mondal2016}D. Mondal, A.K. Pati: Quantum speed limit for mixed states using an experimentally realizable metric. Phys. Lett. A 380, 1395 (2016)
\bibitem{Pires2016}D.P. Pires, M. Cianciaruso, L.C. C\'{e}leri, G. Adesso, D.O. Soares-Pinto: Generalized geometric quantum speed limits. Phys. Rev. X 6, 021031 (2016)
\bibitem{Tomamichel2012}M. Tomamichel, C.C.W. Lim, N. Gisin, R. Renner: Tight finite-key analysis for quantum cryptography. Nat. Commun. 3, 634 (2012)
\bibitem{Coles2014}P.J. Coles, M. Piani: Improved entropic uncertainty relations and information exclusion relations. Phys. Rev. A 89, 022112 (2014)

\bibitem{Horodecki2009}R. Horodecki, P. Horodecki, M. Horodecki, K. Horodecki: Quantum entanglement, Rev. Mod. Phys. 81, 865 (2009)
\bibitem{Nilsen2000}M.A. Nilsen, I.L. Chuang: Quantum Computation and Quantum Information. Cambridge University Press, Cambridge, UK (2000)
\bibitem{Hu2012a}M.L. Hu, H. Fan: Quantum-memory-assisted entropic uncertainty principle, teleportation, and entanglement witness in structured reservoirs. Phys. Rev. A 86, 032338, (2012)
\bibitem{Zou2014}H.M. Zou, H. M. et al.: The quantum entropic uncertainty relation and entanglement witness in the two-atom system coupling with the non-Markovian environments. Phys. Scr. 89, 115101, (2014)

\bibitem{Huang2018a}Z.M. Huang: Dynamics of entropic uncertainty for atoms immersed in thermal fluctuating massless scalar field. Quantum Inf. Process, 17, 73 (2018)
\bibitem{Huang2019}Z.M. Huang, H.Z. Situ: Exploration of entropic uncertainty relation for two accelerating atoms immersed in a bath of electromagnetic field. Quantum Inf. Process. 18, 38 (2019)
\bibitem{Tian2014}Z.H. Tian, J.L. Jing: Dynamics and quantum entanglement of two-level atoms in de Sitter spacetime. Ann. Phys. 350, 1 (2014)
\bibitem{Huang2017a}Z.M. Huang, Z.H. Tian: Dynamics of quantum entanglement in de Sitter spacetime and thermal Minkowski spacetime. Nucl. Phys. B  923, 458 (2017)
\bibitem{Huang2017b}Z.M. Huang: Dynamics of quantum correlation and coherence in de Sitter universe. Quantum Inf. Process. 16, 207 (2017)
\bibitem{Feng2018}J. Feng, X. Huang, Y. Zhang, H. Fan: Bell inequalities violation within non-Bunch¨CDavies states. Phys. Lett. B 786, 403 (2018)
\bibitem{Huang2018b}Z.M. Huang: Protecting quantum Fisher information in curved space-time. Eur. Phys. J. Plus 133, 101 (2018)
\bibitem{Huang2018}X. Huang, J. Feng, Y. Zhang, H. Fan: Quantum estimation in an expanding space-time,  Ann. Phys. 397, 336 (2018)

\bibitem{Vidal2002}G. Vidal, R.F. Werner: Computable measure of entanglement. Phys. Rev. A 65, 032314 (2002)
\bibitem{Peters2004}N.A. Peters, T.C. Wei, P.G. Kwiat: Mixed state sensitivity of several quantum information benchmarks. Phys. Rev. A 70, 052309 (2004)
\bibitem{Sun2010}Q. Sun, M. Al-Amri, L. Davidovich, M. Suhail Zubairy: Reversing entanglement change by a weak measurement. Phys. Rev. A 82, 052323 (2010)

\bibitem{Gorini1976}V. Gorini, A. Kossakowski, and E.C.G. Surdarshan: Completely positive dynamical semigroups of N-level systems. J. Math. Phys.  17, 821 (1976)
\bibitem{Lindblad1976}G. Lindblad: On the generators of quantum dynamical semigroups. Commun. Math. Phys.  48, 119 (1976)
\bibitem{Breuer2002}H.-P. Breuer, F. Petruccione: The Theory of Open Quantum Systems. Oxford University Press, Oxford (2002)

\bibitem{Kubo1957}R. Kubo: Statistical-mechanical theory of irreversible processes. I. General theory and simple applications to magnetic and conduction Problems. J. Phys. Soc. Jpn. 12, 570 (1957)
\bibitem{Martin1959}P.C. Martin, J. Schwinger: Theory of many-particle systems. I. Phys. Rev. 115, 1342 (1959)
\bibitem{Haag1967}R. Haag,  N.M. Hugenholtz, M. Winnink: On the equilibrium states in quantum statistical mechanics.Commun.Math. Phys. 5, 215 (1967)
\bibitem{Barnett1997}S. M. Barnett, P.M. Radmore: Methods in Theoretical Quantum Optics. Oxford University Press, New York (1997)
\bibitem{Bousso2002}R. Bousso, A. Maloney, A. Strominger: Conformal vacua and entropy in de Sitter space. Phys. Rev. D 65, 104039 (2002)
\bibitem{Yu2011}H.W. Yu: Open quantum system approach to the Gibbons-Hawking effect of de Sitter space-time. Phys. Rev. Lett. 106, 061101 (2011)
\end{thebibliography}
\end{document}